\documentclass{ws-procs9x6}  
\def\beq{\begin{equation}}
\def\eeq{\end{equation}}
\def\bea{\begin{eqnarray}}
\def\eea{\end{eqnarray}}
\def\beqa{\begin{equation}\begin{array}{l}}
\def\eeqa{\end{array}\end{equation}}
\def\eqlab#1{\label{eq:#1}}

\def\eref#1{(\ref{eq:#1})}
\def\Eqref#1{Eq.~(\ref{eq:#1})}

\def\lag{{\mathcal L}}

\def\psib{\bar{\psi}}

\def\pa{\partial}
 
\def\slap{p \!\!\!\slash}

\def\half{\mbox{$\frac{1}{2}$}}

\def\quarter{\mbox{$\frac{1}{4}$}}

\def\nn{\nonumber} 

\def\al{\alpha}

\def\ga{\gamma} \def\Ga{{\it\Gamma}}
\def\de{\delta} \def\De{{\it\Delta}}
  \def\eps{\epsilon}

\def\la{\lambda}

\def\pa{\partial}

\begin{document}

\title{Higher spin hadrons
as relativistic fields
\footnote{\uppercase{P}repared for the \uppercase{P}roceedings of 
 \uppercase{NS}tar 2002 \uppercase{W}orkshop, \uppercase{P}ittsburgh, 
\uppercase{PA}, \uppercase{O}ctober 2002.}
}

\author{VLADIMIR PASCALUTSA}

\address{ Department of Physics and Astronomy, Ohio University,
   Athens, OH 45701\\ 
E-mail: vlad@phy.ohiou.edu}


\maketitle

\abstracts{I discuss the problem of consistent interactions
of higher-spin fields and its relevance to resonance physics. 
}

\section{Introduction}
The complexity of QCD does not yet allow us 
to describe low-energy hadronic reactions in terms of 
the underlying quark-gluon dynamics. 
A simpler, albeit more phenomenological,
approach is to seek for a description in terms of hadronic degrees
of freedom.  This approach is usually based on
some form of effective 
Lagrangian (or Hamiltonian) written in terms of hadronic fields, corresponding
to pions, nucleons, $\rho$-mesons, $\Delta$-isobars, etc.

One of the basic problems here arises in the treatment of hadrons with spin
one and higher. This is a very old problem of
{\it consistent interaction of higher-spin fields}.
It first had been addressed in the works of Dirac\cite{Dir36},
Fierz and Pauli\cite{FiP39}, Johnson and Sudarshan\cite{Joh61}, 
Velo and Zwanziger\cite{Vel69}, 
who discovered that by far not any interacting theory
of higher-spin ($s\ge 1$) fields is consistent.  

The problem has to do with the unphysical spin degrees of freedom (DOF)
which are necessarily introduced to achieve a relativistic
formulation of a higher-spin field theory. 
These unphysical DOF are introduced in addition
to the physical $2s+1$ (or, 2) spin DOF which describe the
polarizations of a massive (or, massless) particle with spin $s$.
It turns out a theory is consistent only if
the unphysical DOF {\it decouple}, i.e.,  do not influence
the observables. 
Here I will formulate a consistency condition which
insures the decoupling of unphysical DOF and advocate its
importance in formulating the higher-spin $N^\ast$
couplings.

\section{DOF counting and higher-spin gauge symmetries}
For the sake of manifest Lorentz-invariance
we must operate in terms of tensor and spinor fields.
One of the most common formulations is based on 
symmetric tensors. A boson with spin
$s$ is represented by a rank-$s$ tensor field
$h^{\mu_1\ldots\mu_s}(x)$, while a fermion is described
by a rank $j=s-1/2$ tensor-spinor
$\psi^{(\al)\mu_1\ldots\mu_j}(x)$, symmetrized in
indices $\mu$; index $\al$ is a spinor index.
The question is how to reconcile the number of independent components
of these fields with the number of spin DOF of the corresponding
particle.  For $s\ge 1$ additional constraints
must be imposed on the fields to reduce the number of independent
components to 2 for massless and $2s+1$ for massive situations.
For the massless fields this is done by demanding invariance
of the action under local (gauge) variations of the fields:
\bea
\eqlab{transf}
\de h^{\mu_{1} \cdots\mu_{s}}(x) &=& \pa^{\{\mu_1} \phi^{\mu_2\cdots \mu_s\}}
(x),
\\
\de \psi^{(\al)\mu_{1} \cdots\mu_{j}}(x) &=& \pa^{\{\mu_1} 
\eps^{(\al)\mu_2\cdots \mu_j\}}(x).\nn
\eea
Local symmetries generate constraints and thus it is possible to
formulate free actions with only 2 spin DOF.
The mass term is then introduced such as to (partially) break
the symmetry to raise the number of DOF to $2s+1$.

Rather than going into the DOF counting for arbitrary $s$ (which
can be found in, {\it e.g.}, Ref.~\cite{PaT99}) let us consider
two simplest examples.  A massless spin-1 particle is described
by a vector field $h^\mu$ and the well-known Lagrangian
\beq
\lag_0 = -\quarter F_{\mu\nu}F^{\mu\nu} ,\,\,\,\,\, F^{\mu\nu}
=\pa^{[\mu} h^{\nu]}.
\eeq
Thus, even though the field has 4 components to begin with, the
symmetry of the action under $\de h^\mu =\pa^\mu \phi$ leaves only
2  independent components. The mass term: 
$\lag_m = -\half m^2 h_\mu h^\mu$, raises the spin DOF
number to 3 as is appropriate for a massive spin-1 particle.

A massless spin-3/2 particle is described by a 16-component $\psi_{\mu}$ and
the Rarita-Schwinger Lagrangian (spinor indices omitted):
\beq
 \lag_0 = i\psib_\mu \ga^{\mu\nu\la} \pa_\la \psi_{\nu},
\eeq 
with $\ga^{\mu\nu\la} = \half
(\ga^{\mu}\ga^\nu\ga^\la-\ga^{\la}\ga^\nu\ga^\mu)$.
Here again, due to the symmetry under
$\de\psi_\mu=\pa_\mu\eps$,
(where $\eps$ is a spinor) only 2 components are independent.
The mass term needs to be introduced such that
this number is raised to 4. The following form is known to be uniquely
appropriate:
$\lag_m = - m \,\psib_\mu \ga^{\mu\nu} \psi_\nu $, 
where $\ga^{\mu\nu} = \half [\ga^\mu,\ga^\nu]$.

Free field actions for arbitrary $s$ based on symmetric tensors
were successfully formulated by Singh and Hagen\cite{Sin74} for massive and
by Fronsdal\cite{Fro78} for massless situation.
Interacting theories, on the other hand, appeared to be
much more formidable to formulate consistently. Just giving
a charge to higher-spin particle by a minimal coupling to
electromagnetic field would already lead to serious pathologies,
e.g., negative norm states\cite{Joh61}, solutions to the field equations
propagating faster than light\cite{Vel69}. 
For the spin-3/2 case, all these pathologies can be
related to the fact that the coupling changes the number
constraints leading thus to a theory with wrong DOF content\cite{Pas98}.

This just emphasizes an obvious {\it consistency condition} 
on the interactions
of higher-spin fields: they must {\it 
preserve the DOF counting of the free theory}.
The following statement is less obvious: {\it interactions will be consistent
(preserve the DOF counting) only if they are invariant under the
gauge transformations
similar\footnote{Transformations are {\it similar} if they have 
the same number of transformation parameters and the same order of the
differential operator acting on them. These are the two parameters
which determine the number of constraints imposed by the gauge symmetry.}
to transformations~\eref{transf}.} In other words, interactions
must support the gauge symmetries of the free massless theory, so
that only the mass terms break the symmetries. 

Indeed, since only the breaking of the symmetry
changes the DOF content,  gauge-invariant couplings leave
the DOF counting of the free theory intact, 
for both the massless and massive situations. This argument 
proves at least the sufficiency of the gauge-invariance
requirement. Proof of necessity exists but 
is more involved and will not be presented here. 

The requirement of gauge-invariance of higher-spin couplings
is thus crucial for consistency for {\it both} the massless and
massive fields. It has however been totally ignored in formulating
the couplings of the $N^\ast$ fields. The aim of our investigation
is to correct this situation and implement the consistent couplings
in the $N^\ast$ phenomenology.

\section{Decoupling of the lower-spin DOF}
We will require the $N^\ast$ couplings to be invariant under
the free massless field transformations\footnote{In principle,
symmetry under more general (similar) transformations should be allowed.
However, a nontrivial modification of the transformations, e.g., making
them dependent on other fields, may lead to appearance of new conserved
charges associated with the $N^\ast$ fields. As long as no such charges
are observed in nature one is restricted to  transformations
\eref{transf}.}, \Eqref{transf}.
Corresponding vertex will satisfy the following {\it transversality}
condition:
\beq
\eqlab{tr}
p_{\{\mu_1} \Ga^{\mu_1\cdots \mu_{j}\}} = 0,
\eeq
where $p$ is the momentum and $\mu_{1\ldots j}$
 are the tensor indices corresponding 
to the higher-spin particle; $j=s$ for bosons and $j=s-1/2$ for fermions.

Using this property one can show that there is no coupling
to the lower-spin sector of the propagator. 
Indeed the lower-spin sectors always enter with at least 
one factor of $p^\mu/m$ so when matrix elements 
such as 
$\Ga^{\mu_1\cdots \mu_{j}}\,
 S_{\mu_1\cdots \mu_{j}\nu_1\cdots \nu_{j}}
\,\Ga^{\nu_1\cdots \nu_{j}}$ are computed the net contribution
of the lower-spin sector vanishes. 

To see this effect more explicitly it is helpful 
to write out the propagator in terms of the spin projection
operators. For instance, in the spin-1 case the projector
on pure spin-1 and spin-0 states are:
\beq
{\mathcal P}^{(1)}_{\mu\nu} (p) = g_{\mu\nu}-\frac{p_\mu p_\nu}{p^2}\,,
\,\,\,\,\, {\mathcal P}^{(0)}_{\mu\nu} (p) = \frac{p_\mu p_\nu}{p^2}\,,
\eeq 
and the massive spin-1 propagator is written therefore as
\beq
S_{\mu\nu} (p) =\frac{1}{p^2-m^2} {\mathcal P}^{(1)}_{\mu\nu} (p)
-\frac{1}{m^2}{\mathcal P}^{(0)}_{\mu\nu} (p)\,.
\eeq
Obviously, as long as the vertices involving the spin-1 particle
obey transversality, the spin-0 term drops out of the matrix elements.

In the spin-3/2 case the spin-projection operators are:
\bea
 &&  {\mathcal P}^{(3/2)}_{\mu\nu}(p) 
= g_{\mu\nu} - \frac{1}{3}\gamma_\mu\gamma_\nu
     - \frac{1}{3p^2} (p\hspace{-1.65mm}\slash\gamma_\mu p_\nu
        + p_\mu\gamma_\nu p\hspace{-1.65mm}\slash),\\
 &&   {\mathcal P}^{(1/2)}_{22,\mu\nu}  = 
                 p_\mu p_\nu/p^2,\,\,\,
    {\mathcal P}^{(1/2)}_{12,\mu\nu}  =  p^\varrho p_\nu
                 \ga_{\mu\varrho}/(\sqrt{3}p^2) , \,\,\,
    {\mathcal P}^{(1/2)}_{21,\mu\nu}  =  p_\mu p^\varrho
                 \ga_{\varrho\nu}/(\sqrt{3}p^2) ,\nn
\eea
and the propagator reads:
\beq
   S_{\mu\nu}(p)  =  \frac{1}{\slap-m}
     {\mathcal P}_{\mu\nu}^{(3/2)}-\frac{2}{3m^2}(\slap+m)
     {\mathcal P}^{(1/2)}_{22,\mu\nu} 
      + \frac{1}{\sqrt{3}m} \left({\mathcal P}^{(1/2)}_{12,\mu\nu}
         + {\mathcal P}^{(1/2)}_{21,\mu\nu}\right) 
\eeq
Again, if $p\cdot \Ga=0$ then $\Ga\cdot {\mathcal P}^{(1/2)}\cdot \Ga=0$,
and thus the spin-1/2 sector decouples from the matrix elements.

This decoupling property offers tremendous simplifications
in the treatment of higher-spin hadron fields. When gauge-invariant
couplings are used the lower-spin 
can be dropped from the full relativistic propagators
and only the highest-spin term must be kept\footnote{This cannot be done 
for a coupling that involves the lower-spin sector. Removing
the lower-spin sector by hand in that case violates locality
because of the $1/p^2$ singularity of the projection operators.}.
Certain relativistic
hadron-exchange amplitudes for any $s$ can easily be found,
see, e.g., Appendix of Ref.\cite{PaT00}, where the
$\pi N$ amplitudes are discussed.

\section{Consistent versus conventional couplings}
Most of the commonly used $N^\ast$ couplings are inconsistent
from the DOF-counting point of view presented here, simply because they
do not have the higher-spin gauge symmetry. However,
it can be shown that in perturbation theory the
difference between certain consistent and inconsistent
couplings can be accommodated by specific contact terms\cite{Pas01}.
In other words, when the lower-spin sectors are involved
due to bad couplings, their contribution takes form of contact terms.  
One manages to control, to some extent, the strength of these
contact terms by introducing the so-called off-shell couplings
and thus additional ``off-shell parameters''. The best known
are the off-shell parameters associated with the $\pi N\De$
and $\ga N\De$ couplings, where $\De$ is the spin-3/2
Delta(1232) isobar. 

Often it is claimed that these
off-shell couplings play an important role and are necessary
to describe experimental data. In my view this merely indicates
that some physics is missing but can be reasonably well mimicked
by the contact terms introduced via the off-shell couplings of
the higher-spin $N^\ast$. If these contact terms 
are of unnaturally large importance than an effort needs to
be made to find the explicit mechanisms that it mimics.

A nice example of this situation is provided by two different
calculations of the Compton scattering on the nucleon
in the Delta(1232) resonance region. First calculation, done
by Olaf Scholten and myself\cite{PaS95}, includes
the Delta-excitation using the
conventional $\ga N\De$ couplings $G_1$, $G_2$, 
and two off-shell couplings $z_1$, $z_2$. We found that
the off-shell couplings play a crucial role in reproducing
the data even in relatively low-energy region, around the
pion-production threshold. Second calculation,
by Daniel Phillips and myself\cite{PP}, includes
the Delta using consistent $\ga N\De$ couplings $g_M$ and $g_E$,
as well as chiral one-loop diagrams. In this case there are no off-shell
couplings but the pion loops are included instead. Nevertheless,
the description of the data are of
better quality than in the first calculation. And this is despite
the fact the second calculation has two less free parameters, since
the pion loops in this case bring only the
well-established $m_\pi$, $f_\pi$, and $g_A$.

Certainly without the loops, the calculation with
consistent couplings does much worse phenomenologically
than the conventional one with off-shell couplings. 
So when comparing the performance of  consistent
versus conventional couplings one should keep in mind
that conventional couplings have more free parameters which
adjust the contact term produced by the lower-spin components.
It is in many ways better to separate the discussion
of the genuine $N^\ast$ contribution and the shorter range
effects, which is achieved by using the consistent couplings
and including any necessary contact terms separately.   


\section{Conclusions and outlook}
Relativistic field-theoretic treatment of $N^\ast$ resonances
is needed in approaches based on hadronic degrees of freedom,
such as relativistic potential models, $K$-matrix approach,
chiral perturbation theory. The requirement of physical
spin DOF counting (i.e., $2s+1$ polarization for a massive
particle) constrains the allowed form of the couplings 
of particles with spin higher than one. Such couplings
must support the gauge symmetries of the free {\it massless}
field. At the level of Feynman rules, corresponding
vertices will satisfy the transversality condition, \Eqref{tr}.
It is then easy to see that such gauge-invariant couplings
do not couple to the unphysical lower-spin sector of the field.
These couplings thus allow for a consistent and 
straitforward treatment of the higher-spin $N^\ast$.
It is therefore appears to be promising to implement
these couplings in the $N^\ast$ phenomenology. The work
in this direction is underway.
One of the most challenging aspects of this program ---
the problems of minimal electromagnetic and chiral couplings
of the higher-spin field -- has not been discussed here.
We plan to report on a solution of this problem
in an effective-field-theoretic framework in a nearest future.

\section*{Acknowledgments}
I would like to thank the organizers of this interesting
and stimulating meeting for the warm hospitality and 
generous support.


\begin{thebibliography}{0}
\bibitem{Dir36} P.A.M. Dirac, Proc.~Roy.~Soc.~A {\bf 155}, 447 (1936).
\bibitem{FiP39} M. Fierz and W. Pauli, Proc.~Roy.~Soc.~A {\bf 173}, 211 (1939).
\bibitem{Joh61} K. Johnson and E.C.G. Sudarshan,
                Ann. Phys. (N.Y.) {\bf 13}, 126 (1961).
\bibitem{Vel69} G. Velo and D. Zwanziger,
                Phys. Rev. {\bf 186}, 267, 1337 (1969).
\bibitem{PaT99}
V.~Pascalutsa and R.~Timmermans,
Phys.\ Rev.\ C {\bf 60}, 042201 (1999).
\bibitem{Sin74} L.P.S. Singh and C.R. Hagen,
                Phys. Rev. D {\bf 9}, 898, 910 (1974).
\bibitem{Fro78} C. Fronsdal, Phys. Rev. D {\bf 18}, 3624, 3630(1978);

\bibitem{Pas98}
V.~Pascalutsa,
Phys.\ Rev.\ D {\bf 58}, 096002 (1998).


\bibitem{PaT00}
V.~Pascalutsa and J.~A.~Tjon,
Phys.\ Rev.\ C {\bf 61}, 054003 (2000).

\bibitem{Pas01}
V.~Pascalutsa,
Phys.\ Lett.\ B {\bf 503}, 85 (2001).

\bibitem{PaS95}
V.~Pascalutsa and O.~Scholten,
Nucl.\ Phys.\ A {\bf 591}, 658 (1995).

\bibitem{PP}
V.~Pascalutsa and D.~R.~Phillips,
arXiv:nucl-th/0212024.

\end{thebibliography}
\end{document}